\documentclass{phb-proc4-auth}

\usepackage{graphicx}
\usepackage{amssymb}

\begin{document}
\begin{frontmatter}

%----------------------------------------------------------------------
% Specify destination and version number of the manuscript

\journal{SCES '04}

%----------------------------------------------------------------------
% Title of manuscript

%\title{Quantum Phase Transitions in the Soft-Gap Anderson Model}
\title{Quantum Critical Points in Quantum Impurity Systems}

%----------------------------------------------------------------------
% List of authors
%
% List each author using a separate \author{} command
%
% If there is more than one author address, add a label to each author
% of the form \author[label]{name}.  This label should be identical to
% the corresponding label provided with the \address command.
%
% e.g. if there are three authors from two institutions in USA and 
% France, you can link them to their respective addresses, using
%
% \author[US]{John Doe}
% \author[US,FR]{Jane Doe}
% \author[FR]{Jean Dupont}
% \address[US]{University of Life, Somewhere, USA}
% \address[FR]{Universite de la Vie, Quelque Part, France}
%
% N.B. Unlike the document class used for abstract submissions, it is
% possible to have the author associated with more than one address,
% as shown in the example above.
%

\author{Hyun Jung Lee},
\author{Ralf Bulla\thanksref{DFG}}

%----------------------------------------------------------------------
% List of addresses
%
% If there is more than one address, list each using a separate 
% \address command using a label to link it to the respective author
% as described above
 
\address{Theoretische Physik III, Elektronische Korrelationen und Magnetismus, Universit\"at Augsburg, Germany} 

%----------------------------------------------------------------------
% Title page footnotes
%
% If you need to add qualifying information to any of the authors, 
% use the \thanksref{} command within the \author command.  The 
% argument is the label of a corresponding \thanks[label]{text}
% command which contains the footnote text
%
% e.g. you can acknowledge a funding authority for John Doe, using
%
% \author{John Doe\thanksref{ABC}}
% \thanks[ABC]{This work was supported by Institute of Unphysical 
%    Phenomena under contract no. ABC-123}
%

\thanks[DFG]{This work was supported by the DFG through SFB 484.}

%----------------------------------------------------------------------
% Contact Information
%
% Add the complete postal address, telephone number, fax number, and
% email address of the corresponding author as a special footnote using
% the \corauth[]{} command.  This works in a similar way to the \thanks 
% command.  Add the \corauthref{} command within the \author command.
% The argument is the label of a corresponding \corauth[label]{text}
% command which contains the contact information.  Prefix the text with
% Corresponding Author:
%
% e.g. if the contact author is John Doe,
%
 %\author{John Doe\corauthref{1}}
% \corauth[1]{Corresponding Author: University of Life, 123 Some St.,
%    Somewhere, MI 12345, USA.  Phone: (555) 555-5555 
%    Fax: (555) 555-7777, Email: JDoe@uol.edu}

%\corauth[]{}

%----------------------------------------------------------------------
% Text of abstract

\begin{abstract}
The numerical renormalization group method is used to investigate zero 
temperature phase 
transitions in quantum impurity systems, in particular in 
the soft-gap Anderson model, where an impurity couples to a non-trivial 
fermionic bath. In this case, zero temperature phase transitions occur between two different 
phases whose fixed points can be built up of non-interacting single-particle states. 
However, the quantum critical point cannot be described by non-interacting fermionic or 
bosonic excitations. 
\end{abstract}

%----------------------------------------------------------------------
% Manuscript keywords
%
% Please give two or three keywords in the form: keyword \sep keyword
% e.g. NMR \sep superconductivity
%
% NB The syntax is different from the abstract document class


%----------------------------------------------------------------------
% End of front page

\end{frontmatter}

%----------------------------------------------------------------------
% Manuscript text
%
% Fill in the following space with the manuscript text.
%
% A number of LaTeX commands may be invoked in this space, e.g.
%
% \section{} : to insert a new section title
% \label{}   : to label the numbered section for use in \ref{}
% \cite{}    : to add a reference using the label in \bibitem{}
% 
% A number of LaTeX environments may be used, e.g. 
% \begin{equation}
%     An equation inserted here will be automatically numbered
% \end{equation}  
%
% Please refer to other LaTeX documentation for help on using these
% environments.

Impurity quantum phase transitions have recently attracted
considerable interest (for a review see \cite{BV}). 
These transitions can be observed in systems where
a zero-dimensional boundary (the impurity) interacts with a 
bath of free fermions or bosons. Examples include one or
two magnetic impurities coupling to one or more fermionic baths \cite{BV}
and the spin-boson model, where a two-level system couples to a bosonic
bath \cite{Leggett,BTV}

A very well studied model is the soft-gap Anderson model \cite{withoff}:
\begin{eqnarray}
     H &=&   \varepsilon_{f} \sum_{\sigma} f^\dagger_{\sigma} f_{\sigma}
          + U f^\dagger_{\uparrow} f_{\uparrow}
                              f^\dagger_{\downarrow} f_{\downarrow}
   \nonumber  \\ 
      & & + \sum_{k\sigma} \varepsilon_k c^\dagger_{k\sigma} c_{k\sigma}
       + V \sum_{k\sigma} \big( f^\dagger_{\sigma} c_{k\sigma} +
               c^\dagger_{k\sigma}  f_{\sigma} \big)  \ .
\label{eq:model}
\end{eqnarray}
(Notations are standard.) This model has the same form as the
single impurity Anderson model but for the soft-gap model we require that
the hybridization function 
$\Delta(\omega)=\pi V^2\sum_k \delta(\omega-\varepsilon_k)$ has
a soft-gap at the Fermi level, $\Delta(\omega)=\Delta_0 \vert\omega\vert^r$,
with an exponent $r>0$.

The model eq.~(\ref{eq:model}) shows a non-trivial 
zero-temperature quantum phase transition 
between a local moment phase (LM) and a strong coupling phase (SC)
at a {\em finite} value of the 
coupling $\Delta_0$. Since the first approach using a large degeneracy 
technique \cite{withoff}, intensive studies have been performed to clarify 
various properties of 
this model, such as the impurity susceptibility, specific heat and entropy 
\cite{ingersent,bulla}, and dynamic quantities \cite{LMA}. 
An important issue, which is far from being fully understood, is the characterization
of the quantum critical points. To this end, we employ the numerical renormalization
group method suitably extended to handle 
non-constant couplings $\Delta(\omega)$ 
\cite{ingersent,bulla}. This method allows a non-perturbative calculation of
the many-particle spectrum and physical properties in the whole 
parameter regime
of the model eq.~(\ref{eq:model}), in particular in the low-temperature limit, so that the structure
of the quantum critical points is accessible.

 \begin{figure}[htb]
     \centering
     \includegraphics[width=0.3\textwidth,angle=-90]{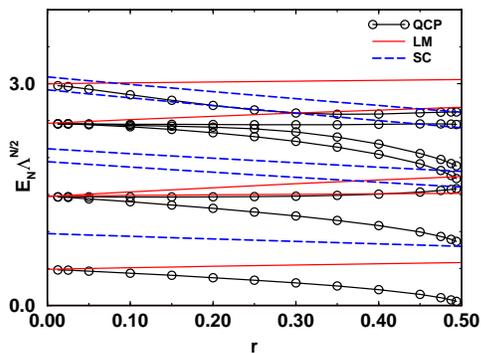}
     \caption{Dependence of the many-particle spectra for the three
            fixed points of the symmetric soft-gap Anderson model on
            the exponent $r$.} 
     \label{fig1}
  \end{figure}

In  Fig.~\ref{fig1}, 
the many-particle
spectra corresponding to the local moment (dotted lines), 
the strong coupling (dashed lines), and the quantum critical (QCP, lines with 
circles) 
fixed points of the symmetric soft-gap model
are shown as functions of the exponent $r$
(for a similar figure, see Fig.~13 in \cite{ingersent}). 
Each set of fixed points is extracted from the lowest six 
levels with quantum-numbers $Q=-1$, $S=0$.

The fixed-point structure of the strong coupling 
and the local moment phase  can be easily understood as that of a 
free conduction electron 
chain. The combination of the single-particle states of the free 
chain leads to the 
degeneracies seen in Fig.~\ref{fig3} below.

 \begin{figure}[htb]
     \centering
     \includegraphics[width=0.4\textwidth]{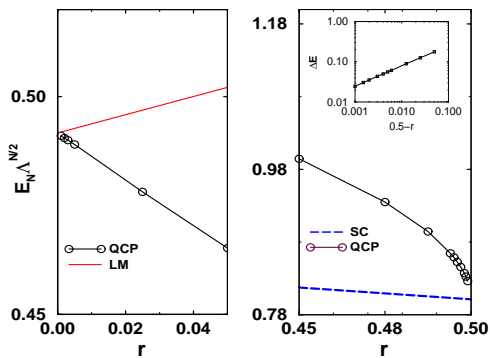}
     \caption{Same data as in Fig.~\ref{fig1} for the region close
            to $r=0$ (left panel) and $r=0.5$ (right panel). The
            deviation between QCP and LM is linear for small $r$
            while the deviation between QCP and SC is proportional
            to $\sqrt{0.5-r}$, as can be seen in the inset of
            the right panel.} 
     \label{fig2}
\end{figure}

What information can be extracted from such a level structure? First we observe
that the levels of the QCP approach the levels of LM(SC) in the limits 
$r\to0$ ($r\to0.5$). The limiting behaviour is illustrated in 
Fig.~\ref{fig2}: the QCP levels
deviate linearly from the LM levels for $r\to0$ 
while the deviation $\Delta E$
 to the SC levels is proportional to $\sqrt{0.5-r}$
for $r\to0.5$. This has direct consequences for physical properties at the QCP;
the local susceptibility at the QCP, for example, shows a square-root
dependence on $(0.5-r)$ close to $r=0.5$ \cite{ingersent}.

A plot similar to Fig.~\ref{fig1} can be calculated for the
spin-boson model, using the numerical renormalization group
method as in Ref.~\cite{BTV} (data not shown here). In the
spin-boson model, the many-particle levels of the QCP turn out
to approach the levels of the delocalized (localized) fixed point in the
limit $s\to0$ ($s\to1$), with $s$ the exponent of the bath spectral
function $J(\omega)\propto\omega^s$.

It would be nice to compare these numerical results with a
perturbative expansion around the limits $r=0$ and $r=0.5$. 
Such an expansion has
been performed for small $r$ in Ref.~\cite{KV}. Interestingly,
the linear deviation between LM and QCP levels for small $r$ cancels
in the evaluation of the impurity contribution to the entropy, and
the correction turns out to be $\propto r^3$. The level structure
of the QCP 
itself might be accessible to perturbational approaches, at least
in the limits $r=0$ and $r=0.5$; the deviations would then appear
as marginal perturbations of the LM and SC fixed points.

 \begin{figure}[htb]
     \centering
     \includegraphics[width=0.29\textwidth,angle=-90]{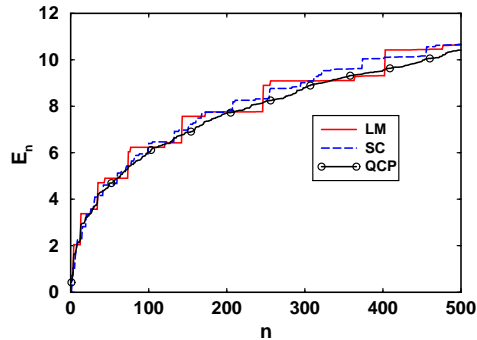}
     \caption{Many-particle energies of the three fixed points
            of the symmetric soft-gap model versus the label $n$
            for $r=0.3$.} 
     \label{fig3}
\end{figure}

An alternative way to characterize the spectra of the various fixed
points is shown in Fig.~\ref{fig3} 
where we plot the many-particle energies $E_n$
versus the label $n$ which simply counts the many-particle states starting
from the ground-state $n=1$. Degeneracies due to symmetries of the model
are not considered. We clearly see that the LM and SC fixed points show
additional degeneracies as they are built up from single-particle excitations;
this gives rise to the plateaus in $E_n$.

In the case of the quantum critical fixed points, degeneracies due to the 
combination of single-particle levels are missing,
which indicates that the 
quantum 
critical point cannot be constructed from
non-interacting single-particle states.

%----------------------------------------------------------------------
% Reference section
%
% List each reference with a separate \bibitem{} command.  The
% argument contains the label that is used in the \cite{} command
% in the main text
%
% e.g.
%
%    This follows our pioneering work on TdB2\cite{TdB2}.
%
% \bibitem{TdB2}
% J. Doe, J. Doe, and J. Dupont, J. Irrep. Res. 10 (2000) 1000.

%----------------------------------------------------------------------
% Figures and Tables
%
% Insert figures and tables at the end of the document unless you
% are familiar with the LaTeX positional options.
%
% \begin{figure}
%     \centering
%     \includegraphics{filename.eps}
%     \caption{Insert figure caption here} 
% \end{figure}  
%
% \begin{table}
%     \centering
%     \begin{tabular}
%     Insert table here
%     \end{tabular}
%     \caption{Insert table caption here}
% \end{table}  

%\newpage
% \begin{figure}[htb]
%     \centering
%     \includegraphics[width=0.60\textwidth]{r_limit_xfig.eps}
%     \caption{Fig.2} 
% \end{figure}  

%
% Please refer to other LaTeX documentation for help on using these
% environments.

%----------------------------------------------------------------------
% Terminate document


\begin{thebibliography}{00}

\bibitem{BV} R. Bulla and M. Vojta,
   in {\it Concepts in Electron Correlations},
   A.C. Hewson and V. Zlatic (eds.), Kluwer Academic Publishers, Dordrecht (2003), 209. 
\bibitem{Leggett}
%A.~J. Leggett {\em et al.},
 A.~J. Leggett, S. Chakravarty, A.T. Dorsey, M.P.A. Fisher, A. Garg, and W. Zwerger,
Rev. Mod. Phys. {\bf 59}, 1 (1987).

\bibitem{BTV} R. Bulla, N. Tong, and M. Vojta,
              Phys. Rev. Lett. {\bf 91}, 170601 (2003). 
\bibitem{withoff} D. Withoff and E. Fradkin, Phys. Rev. Lett. {\bf 64}, 1835 (1990).
\bibitem{ingersent} C. Gonzalez-Buxton and K. Ingersent, Phys. Rev. B {\bf 57}, 14254 (1998).
\bibitem{bulla} R. Bulla, T. Pruschke and A.C. Hewson, J. Phys.: Condens. Matter {\bf 9}, 10463 (1997).
\bibitem{LMA}  M.T. Glossop and D.E. Logan,
               J. Phys.: Condens. Matter {\bf 15} (2003) 7519. 
\bibitem{KV}  M. Kircan and M. Vojta,
              Phys. Rev. B {\bf 69}, 174421 (2004).

\end{thebibliography}
\end{document}